\documentclass[prb,twocolumn,superscriptaddress,aps,longbibliography]{revtex4-1}
\usepackage[utf8]{inputenc}

\usepackage{graphicx}
\usepackage{bm}
\usepackage{amsmath}
\usepackage{amssymb}
\usepackage{float}
\usepackage{xcolor}
\usepackage[colorlinks=true,citecolor=blue]{hyperref}

\DeclareGraphicsExtensions{.png,.jpg,.eps}

\renewcommand{\r} {\mathbf{r}} 
\newcommand{\q} {\mathbf{q}} 
\renewcommand{\H} {\mathcal{H}} 

\newcommand{\gap} {\mathcal{G}}
\begin{document}

\author{Maryam Khosravian}
\affiliation{Department of Applied Physics, Aalto University, 02150 Espoo, Finland}

\author{Jose L. Lado}
\affiliation{Department of Applied Physics, Aalto University, 02150 Espoo, Finland}

\title{Impurity-induced excitations in a topological two-dimensional ferromagnet/superconductor van der Waals moiré heterostructure}

\begin{abstract}
The emergence of a topological superconducting state in
van der Waals heterostructures provides a new platform for
exploring novel strategies to control topological superconductors.
In particular, impurities in van der Waals heterostructures, generically featuring a moire pattern,
can potentially lead
to the unique interplay between atomic and moire length scales, a feature
absent in generic topological superconductors.
Here we address the impact of non-magnetic
impurities on a topological moire superconductor,
both in the weak and strong regime, considering both periodic arrays and single impurities
in otherwise pristine infinite moire systems.
We demonstrate a fine interplay between impurity induced modes and the moire length,
leading to radically different spectral and topological properties
depending on the relative impurity location and moire lengths.
Our results highlight the key role of impurities in van der Waals heterostructures
featuring moire patterns, revealing the key
interplay between length and 
energies scales in artificial moire systems.

\end{abstract}
\date{\today}
\maketitle

\section{Introduction}
The design of artificial topological superconductors is one of the most active areas in designer quantum materials\cite{Beenakker2013,PhysRevLett.111.056402,Alicea2011,Feldman2016,NadjPerge2014,PhysRevB.88.155420,PhysRevB.96.174521,PhysRevLett.114.236803,PhysRevLett.102.216404,PhysRevLett.108.096802,PhysRevLett.111.186805,PhysRevLett.121.037002,PhysRevResearch.3.033049,PhysRevX.5.041042,PhysRevLett.114.236803,RevModPhys.83.1057}, 
fueled by their fundamental interest and their potential for future topological computing architectures\cite{Alicea2012,PhysRevX.6.031016,PhysRevLett.104.040502,PhysRevLett.105.177002}.
Engineering topological superconductivity requires three different ingredients
magnetism, spin-orbit coupling, and superconductivity, 
features that have been harvested in a variety of platforms, including semiconducting
nanowires\cite{Mourik2012,Deng2016,Das2012},
atomic chains\cite{NadjPerge2014,Schneider2021}, topological insulators\cite{Lpke2020,Jck2019},
and van der Waals materials\cite{Kezilebieke2020,Kezilebieke2021,Kezilebieke2022}.
In particular,  besides all the rich physics of topological superconductors,
van der Waals topological superconductors such as
CrBr$_3$/NbSe$_{2}$ heterostructures\cite{Kezilebieke2020,Kezilebieke2021,Kezilebieke2022},
display a unique feature stemming from their van der Waals nature\cite{Geim2013,Liu2016},
the emergence of a moire pattern.

\begin{figure}[t!]
    \centering
    \includegraphics[width=\columnwidth]{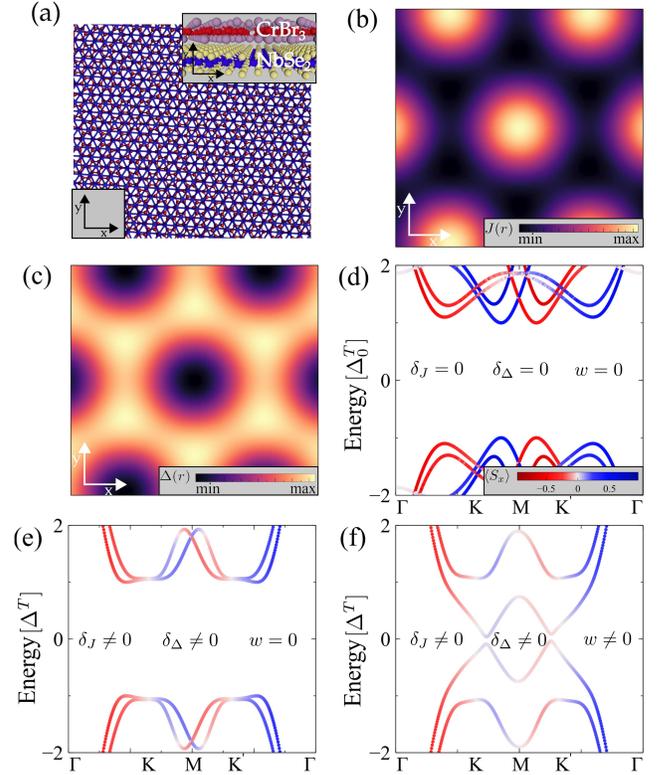}
    \caption{(a) Schematic view of an artificial 
    CrBr$_3$/NbSe$_2$ moire topological superconductor. 
    Panels (b,c) show the real-space
    modulation of exchange coupling $J(\mathbf{r})$ (b)
    and superconductivity $\Delta(\mathbf{r})$ (c) in the moire unit cell. 
    Panels (d,e,f) 
    electronic band structure of a (d) uniform, (e) moire, 
    and (f) moire with an impurity topological
    superconductor.     We used
    $J_{0}$=2$\Delta_{0}$, $\lambda$ =2$\Delta_{0}$, $\mu=3t$, $\delta_{J} = 2J_{0}$, $\delta_\Delta=1.4\Delta_{0}$.
    }
    \label{fig:fig1}
\end{figure}

Artificial van der Waals topological superconductors,
and moire heterostructures in general,
display two length scales, the original lattice
constant and the emergent moire length\cite{PhysRevLett.99.256802,PhysRevB.82.121407,Bistritzer2011,PhysRevB.93.235153,PhysRevB.92.075402}.
The existence of a moire pattern leads to a superconducting state with an associated
moire electronic structure, and more importantly, 
a spatially modulated
structure directly inherited from
the moire pattern\cite{Song2021,Andersen2021}.
In particular, atomic defects in two-dimensional materials
including substitutional
elements and vacancies have a relevant length scale
stemming from the microscopic lattice constant\cite{PhysRevLett.104.036802,PhysRevLett.104.096804,Nguyen2017,Wang2018,Barja2019}, 
and therefore can give rise to a rich interplay with the moire length.
While impurities in uniform topological superconductors
have been widely studied\cite{RevModPhys.78.373,PhysRevB.91.201411,PhysRevB.88.155435,PhysRevB.93.035134,PhysRevB.93.014517,Kaladzhyan2016,PhysRevB.92.174514},
the interplay between local impurities in moire systems has
remained relatively unexplored\cite{Brihuega2017,Ramires2019,PhysRevResearch.2.033357,PhysRevMaterials.3.084003}.

In this manuscript, we address the impact of non-magnetic impurities on 
artificial moire topological superconductors.
We show that interplay between the moire pattern and local impurities gives rise 
to radically different impacts of the defects depending on the specific location.
Our manuscript is organized as follows. First, in section
\ref{sec:model} we introduce the model for a moire topological superconductor.
In Sec.~\ref{sec:strong}
we address the impact of strong impurities
in the topological state.
In Sec.~\ref{sec:weak} we address the impact of weak impurities
in the moire superconductor.
In Sec.~\ref{sec:amp}
we study the interplay between the amplitude of the moire pattern
and the location of the impurity.
In Sec.~\ref{sec:single} we address the emergence of ingap modes for single impurities
in otherwise pristine systems. 
In Sec.~\ref{sec:edge} we examine the emergence
of topological moire edge modes
in the defective and pristine topological moire superconductor.
Finally, in Sec.~\ref{sec:summ} we summarize our conclusions.

\section{Model}
\label{sec:model}

Here we will consider a generic topological superconductor realized in a van der Waals
heterostructure, as realized in a CrBr$_3$/NbSe$_2$ heterostructure\cite{Kezilebieke2020}.
In this material, the relative lattice mismatch and rotation between the
CrBr$_3$ ferromagnet and the NbSe$_2$ superconductor
leads to the emergence of a moire pattern,
as shown in Fig. \ref{fig:fig1}a.
In particular, 
moire pattern is expected to directly impact
underlying Hamiltonian of the system due to the
local structural modulation\cite{PhysRevB.104.195156,PhysRevB.104.075126,PhysRevLett.124.136403,PhysRevB.103.085109,Karpiak2019}.
Such a modulation directly imprints Yu-Shiba-Rusinov states following the moire pattern\cite{Kezilebieke2022}.
Due to the structural
modulation, including the
local hoppings\cite{PhysRevResearch.2.043127}, induced spin-orbit coupling\cite{PhysRevB.104.195156},
chemical potential\cite{Zhang2020}, exchange field\cite{Ubrig2019} 
and superconducting proximity\cite{Trainer2020} will be modulated. For the
sake of concreteness, here we will focus on the two parameters
whose modulation is expected to be most sizable,
the local superconducting order
and the proximity-induced exchange field.
It is worth noting that, beyond the currently realized
CrBr$_3$/NbSe$_2$ heterostructure
displaying topological superconductivity\cite{Kezilebieke2020,Kezilebieke2021,Kezilebieke2022}, 
a variety of other artificial van der Waals
systems can potentially lead to topological superconductivity. As two-dimensional
ferromagnets, materials such as CrBr$_3$\cite{Kim2019}, CrI$_3$\cite{Huang2017} and
CrBr$_{3-x}$I$_x$\cite{Tartaglia2020},
provide potential out-of-plane ferromagnetic monolayers, 
whereas NbSe$_2$\cite{Ugeda2015}, NbS$_2$\cite{Yan2019}, TaS$_2$\cite{NavarroMoratalla2016} and TaSe$_2$\cite{Lian2019}
and their alloys\cite{Zhao2019} would provide
van der Waals superconductors. For all the combinations between
ferromagnets and superconductors above, the resulting heterostructure
will show a moire pattern between a honeycomb ferromagnet and a
triangular superconductor, displaying an approximate C$_3$ rotational
symmetry. 

With the previous platforms in mind, we now turn to address a
minimal effective model for the previous heterostructures.
While the specific parameters of the model for each material combination
should be estimated using first-principles calculations\cite{PhysRevB.104.195156,PhysRevB.104.075126,PhysRevLett.124.136403,PhysRevB.103.085109,Karpiak2019}, here we will
focus on addressing the universal features that arise due the interplay
of the moire superconductor and the local impurities\cite{Brihuega2017,Ramires2019,PhysRevResearch.2.033357,PhysRevMaterials.3.084003}.
The structure moire pattern directly gives rise
to a modulation in the induced exchange coupling and an s-wave
superconductivity, whose
spatial profiles are shown in Fig.~\ref{fig:fig1}bc.
The electronic structure of the heterostructure
is modeled with an atomistic Wannier orbital
per Nb site forming a triangular lattice,
where the moire pattern is incorporated in the
modulation of the Hamiltonian parameters.
The full Hamiltonian takes the form

\begin{equation}
\H_{0} = 
\H_{\text{kin}}
+
\H_J +
\H_{R}  +
\H_{\text{SC}}
\label{eq:h1}
\end{equation}
with
\begin{equation}
\H_{\text{kin}} = 
t \sum_{\langle ij\rangle,s} c^{\dagger}_{i, s}c_{j,s} +\sum_{i}\mu({\r}) c^{\dagger}_{i, s}c_{i,s}
\end{equation}
where $c_{n,s}^{\dagger}$( $c_{n,s}$) denotes the creation (annihilation) fermionic operator 
with spin $s$ in site $n$. $t$ is the hopping parameter and $\langle i, j\rangle $ runs
over nearest neighbors, and $\mu$ is the chemical potential. 
The term

\begin{equation}
    \H_J = 
    \sum_{i,s,s'} J(\r) \sigma_z^{s,s'} c^{\dagger}_{i,s}c_{i,s'} 
    \label{ej}
\end{equation}

is the exchange coupling induced by the underlying ferromagnet, obtained
by integrating out the degrees of freedom of the magnetic monolayer. The term

\begin{equation}
\H_{R} = 
    i\lambda  \sum_{\langle ij \rangle, s s^{'}} 
    \mathbf{d}_{ij} 
    \cdot 
    \mathbf{\sigma} ^{s,s^{'}} c^{\dagger}_{i, s}c_{j,s^{'}}
\end{equation}

is the Rashba spin-orbit coupling
arising due to the
broken mirror symmetry at the
NbSe$_2$/CrBr$_3$ interface,
$\mathbf \sigma$ are the spin Pauli
matrices, $\lambda$ controls the spin-orbit coupling constant
and
$\mathbf{d}_{ij} = \mathbf{r}_i - \mathbf{r}_j$.
The term

\begin{equation}
\H_{\text{SC}} = 
    \sum_{i} \Delta (\r) c^{\dagger}_{i,\uparrow}c^{\dagger}_{i,\downarrow}+h.c.
   \label{esc}
\end{equation}

is the s-wave superconducting order.  
 $J(\mathbf{r})$ and $\Delta(\mathbf{r})$ parameterize the exchange coupling and 
induced s-wave superconductivity. 

Local non-magnetic impurities are included
adding a potential scattering term
of the form

\begin{equation}
    \H_{\text{imp}} = 
    w \sum_s c^{\dagger}_{n,s}c_{n,s}
\end{equation}

where $\H_{\text{imp}}$ defines the impurity Hamiltonian at site $n$ with an on-site 
potential $w$. The full Hamiltonian of the defective
system takes the form

\begin{equation}
    \H = \H_0 + \H_{\text{imp}}
\end{equation}

As noted above, 
the moire profile in our effective model arises from the combination of modulated exchange
coupling and s-wave
superconductivity\cite{PhysRevResearch.3.013262}. 
The interplay between the exchange field and the superconducting
order, as two competing orders, results in an opposite modulated
superconductivity and therefore exhibits a modulated moire pattern in the whole platform. 
We account for this
by defining a potential in real space with the functional form of

\begin{equation}
f(\r) = c_0 + c_1 \sum _{n=1}^{3}\cos({\it {R^{n}}\mathbf{\q} \cdot \r}) 
\label{eq:modul}
\end{equation}

where ${\q}$ is the moire superlattice wave vector, 
and $R_{n}$ is the rotation matrix which
conserve $C_3$ symmetry
and $c_0,c_1$ are chosen
so that $f(\r)\in (0,1)$. The spatial 
profiles $J(\mathbf{r})$ and $\Delta(\mathbf{r})$ 
are written in terms
of the previous spatial dependence as

\begin{align}
J(\r) = J_{0}+ {\chi} \delta_{J} f(\r) \nonumber \\
\Delta(\r) = \Delta_{0}+ {\chi} \delta_{\Delta} (1-f(\r))
\label{eq:amplitude}
\end{align}

$J_{0}$ and $\Delta_{0}$ controls
the average magnitude of the modulated exchange and superconducting profiles,
whereas $\delta_{J}$ and $\delta_{\Delta}$ control
the amplitude of the moire modulation, respectively. 
We introduce $\chi \in (0,1)$ as a parameter which allows us to adiabatically switch
between a uniform or moire system, and here it is taken to be 1. As noted above, the relative signs of $f(\r)$ in $J(\r)$ and $\Delta(\r)$ are taken so that
when the exchange is maximum, the local superconducting order is minimum.
We now elaborate on the mechanism that yields those
two parameters as the dominating modulations.
The stacking heavily influences the value of the exchange, and ultimately
it can also impact its sign\cite{Soriano2019,Sivadas2018}. This strong modulation of the exchange directly affects the local
superconducting order, as a finite
exchange field locally quenches the s-wave pairing.
Therefore the superconducting and exchange modulations are anticorrelated. 
This is the behavior directly
expected from the pair-breaking
effect of the exchange field in an s-wave superconductor\cite{DeGennes2018,RevModPhys.77.935},
and arises naturally from a selfconsistent treatment of the superconducting state
in the presence of the moire modulated exchange.
Experimentally, the impact of moire modulations in NbSe$_2$/CrBr$_3$ has been directly
observed by imaging the spatial dependence of the Yu-Shiba-Rusinov (YSR)
bands at energies inside the gap, directly reflecting the modulation of the exchange coupling\cite{Kezilebieke2022}.
We note that, for the superconductor taken as NbSe$_2$, the closest saddle point to the Fermi energy
that will have the strongest moire effect is located at the M point\cite{PhysRevB.92.134510}.
Finally, of course the moire can also influence the other parameters, but their effect is expected to be
be substantially smaller in comparison with the exchange field.

We now briefly elaborate in the procedure to solve the previous Hamiltonian.We take as starting point the effective Hamiltonian for a periodic moire supercell,
that takes the form
\begin{equation}
\label{eq:bdg}
\mathcal{H} =  \sum 
\Gamma^{\alpha,\beta}_{i,j,s,s'} c_{i,\alpha,s}^{\dagger} c_{j,\beta,s'} + \sum (\Delta_{\alpha} c_{i,\alpha,\uparrow}^{\dagger}c_{i,\alpha,\downarrow}^{\dagger} + h.c.)
\end{equation}
where $c^{\dagger}_{i,\alpha,s}$ denotes the creation operator at unit cell $i$, site $\alpha$ and spin $s$
and $\Gamma^{\alpha,\beta}_{i,j,s,s'}$ contains the hopping, spin-orbit coupling
and exchange proximity effects.
The previous Hamiltonian, periodic in the supercell,
can be diagonalized by defining the Bloch's operators
\begin{eqnarray}
    c_{j,\alpha,s} = \sum_{\mathbf{k}}e ^{i\mathbf{k}\cdot{\mathbf{R_{j}}} } c_{\mathbf{k},\alpha,s} 
\end{eqnarray}
leading to the Hamiltonian in momentum space

\begin{equation}
\label{eq:aprox}
   \mathcal{H} = 
   \sum
   \Gamma^{\alpha,\beta}_{s,s'} (\mathbf{k})
   c_{\mathbf{k},\alpha,{s}}^\dagger c_{\mathbf{k},\beta,{s'}}+
   \sum
   [\Delta_{\alpha} c_{{\mathbf{k},\alpha, \uparrow}}^\dagger c_{-\mathbf{k},\alpha,{\downarrow}}^\dagger + h.c.] 
\end{equation}
where $\Gamma^{\alpha,\beta}_{s,s'} (\mathbf{k})$ is the Fourier transform in the unit cell indexes $i,j$ of the matrices
$\Gamma^{\alpha,\beta}_{i,j,s,s'}$.
To diagonalize the previous Hamiltonian, we define a new fermionic operators
$c^\dagger_{\mathbf{k},s} \equiv (c^\dagger_{\mathbf{k},1,s},...,c^\dagger_{\mathbf{k},n,s})$ and 
$\Psi^\dagger_\mathbf{k} = 
(c^\dagger_{\mathbf{k},\uparrow},
c^\dagger_{\mathbf{k},\downarrow},
c_{\mathbf{-k},\downarrow},
c_{\mathbf{-k},\uparrow})
$. The Hamiltonian in this basis can be written as

\begin{equation}
    \mathcal{H} = \frac{1}{2}\sum
    \Psi^\dagger_\mathbf{k} H_{BdG} \Psi_\mathbf{k}
\end{equation}
where 
the Bogoliubov de Gennes (BdG) Hamiltonian 
$H_{BdG}$
is diagonalized in terms of the new operators,
leading to a diagonal form
\begin{eqnarray}\label{eq:quadratic}
\mathcal{H} = 
\frac{1}{2}
\sum
\epsilon_{\mathbf{k},\alpha} 
\Psi^\dagger_{\mathbf{k},\alpha} \Psi_{\mathbf{k},\alpha}
\end{eqnarray}
where $\epsilon_{\mathbf{k},\alpha} $ are the
BdG eigenvalues and 
$\Psi^\dagger_{\mathbf{k},\alpha}$ the BdG eigenstates.

In practice, the calculation of the electronic structure for a moire supercell
requires diagonalizing a $4n\times4n$ matrix, with $n$ the number of sites per
supercell. To compute the electronic structure in a ribbon geometry, an analogous procedure
as the one outline above is carried out for the supercell of the nanoribbon, that contains
several moire unit cells. Finally, to compute surface spectral functions, the
embedding formalism described later 
in the manuscript is directly applied to the matrices defined
by the BdG Hamiltonian.

It is instructive to first look at the electronic structure and topological character of
the uniform system of Eq.~\ref{eq:h1} when moire potential is switched off. For the Hamiltonian parameters we choose the following setup: $J_{0}$=2$\Delta_{0}$, $\lambda$ =2$\Delta_{0}$, and the chemical potential set to $\mu=3t$. In the uniform limit, for a supercell of size
$9\times 9$ as shown in Fig.~\ref{fig:fig1}d, the
system shows a finite gap of topological character,
where the energy is measured in terms of the topological gap of the uniform system
$\Delta^T_0$. 
This topological superconducting gap is obtained by taking
appropriate ratios of the exchange field, Rashba spin-orbit coupling, 
superconducting order and chemical potential,
and features a topological gap with Chern number $C=2$\cite{Kezilebieke2020}.

Keeping the same average values of the exchange, superconductivity,
Rashba spin-orbit coupling and chemical potential, we now move on to the moire system, switching on the exchange and superconducting modulation $\delta_J$ and $\delta_\Delta$, which are set to $\delta_{J} $= 2$J_{0}$, $\delta_\Delta$=1.4$\Delta_{0}$.
The electronic structure of the modulated system is shown in 
Fig.~\ref{fig:fig1}e, and features a topological
gap with Chern number $C=2$,
where the energy is measured in terms of the topological gap of the moire
system $\Delta^T$. 
We note that
there is not a simple analytic relation between both gaps, and in our calculations
we explicitly compute both.
Phenomenologically, in the regime
we will target
we find that the moire modulation slightly decreases
the topological gap, 
in the worst case scenario by up to
a factor three in comparison with the uniform case.
In this last moire superconductor, we now consider the impact of a single
strong non-magnetic impurity per moire unit cell with $w=2t$.
The electronic structure of the moire modulated system with a single impurity is
shown in Fig.~\ref{fig:fig1}f. 
It is clearly observed that the gap gets drastically reduced in comparison
with the moire pristine limit of Fig.~\ref{fig:fig1}e.
The dramatic impact of the local impurity is a consequence
of the unconventional nature of the topological
superconducting gap of the moire heterostructure.
While this feature also appears in generic artificial
topological superconductors\cite{RevModPhys.78.373,PhysRevB.91.201411,PhysRevB.88.155435,PhysRevB.93.035134,PhysRevB.93.014517,Kaladzhyan2016,PhysRevB.92.174514}, 
the existence of the moire pattern
gives rise to a complex interplay between the local impurity and 
the moire length as we address in the next sections.

\section{Strong impurities in topological moire superconductors}
\label{sec:strong}

\begin{figure}[t!]
    \centering
   \includegraphics[width=\columnwidth]{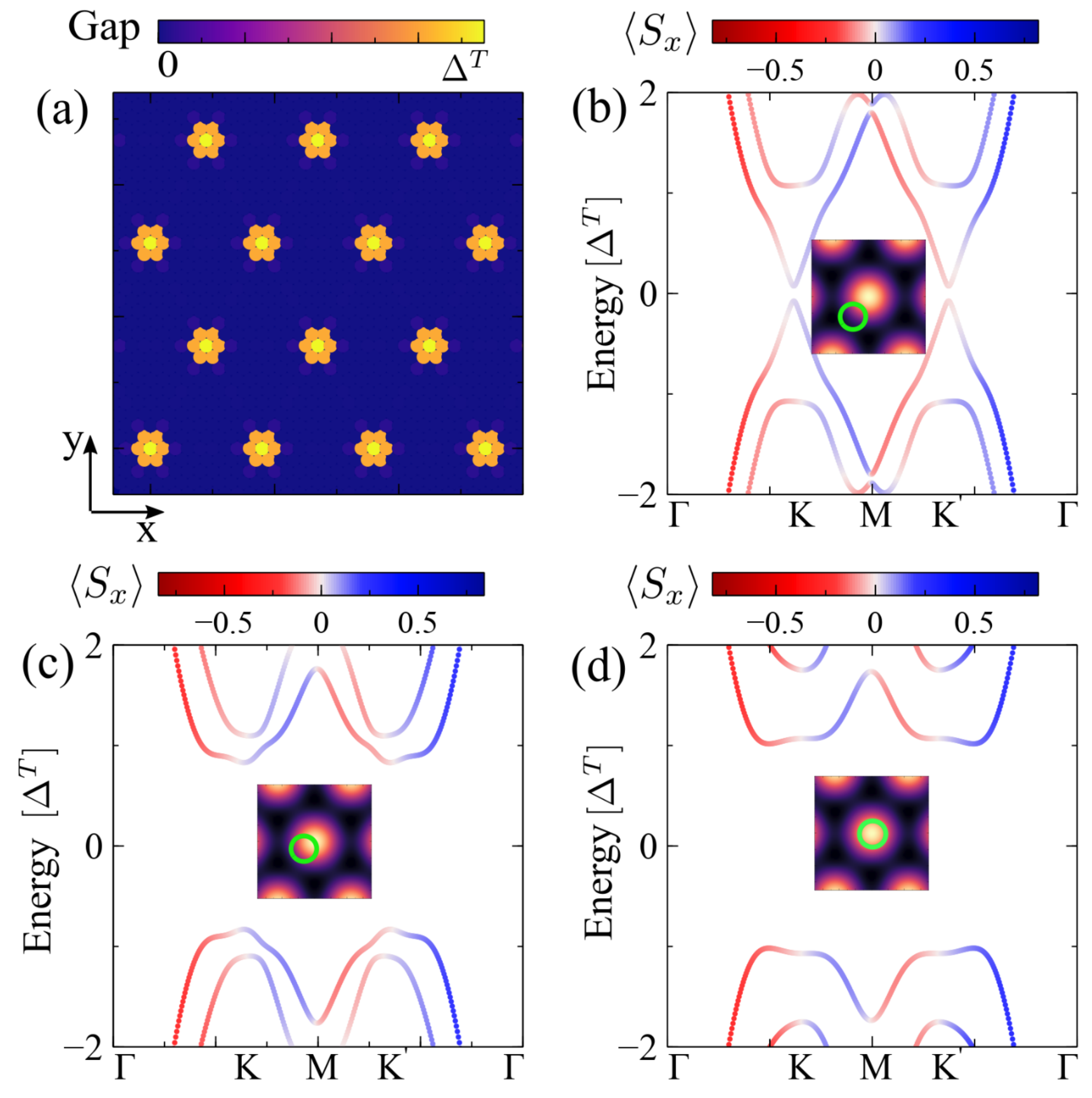}
    \caption{(a) Gap as a function of impurity location in the strong
    impurity regime $w=2t$. Panels (bcd) show the electronic structure
    for three impurity locations, showing that halfway between exchange maxima, the gap is minimal
    (b). In comparison, close to the exchange maxima, the gap
    remains nearly unchanged (c,d).
    We used
    $J_{0}$=2$\Delta_{0}$, $\lambda$ =2$\Delta_{0}$, $\mu=3t$, $\delta_{J} = 2J_{0}$, $\delta_\Delta=1.4\Delta_{0}$.
    }
    \label{fig:strong}
\end{figure}

We now examine in detail
the case of a strong impurity potential. Strong impurities
in the effective model are associated to chemical impurities\cite{Lin2016,Wang2018,PhysRevB.92.235408,MahjouriSamani2016}
and vacancies\cite{Pandey2016} in the
dichalcogendie superconductor, and 
give rise to a strong scattering center. In particular,
for the dichalcogenide superconductor,
chemical impurities such as oxygen\cite{Barja2019} will give
rise to a strong disruption
of the electronic structure. 
Substitutional oxygen impurities, in particular, are expected to create a
strong local scattering, comparable and even higher than the hopping of the
low energy Wannier orbitals\cite{PhysRevB.92.235408}.
In contrast, atomic replacements such as substitutional
S atoms in NbSe$_2$ and TaSe$_2$ would give rise to weaker
scattering centers\cite{Li2017,Zhao2019}.

It is worth noting that, since the scattering potential induced by
the previous impurities
is non-magnetic, the emergence of in-gap modes stems from the non-trivial
nature of the superconducting gap in the moire system\cite{RevModPhys.78.373}. In particular,
non-magnetic impurities in conventional s-wave superconductors are well known
to not give rise to in-gap states as given by Anderson's theorem\cite{Anderson1959,PhysRevB.94.104501,PhysRevB.98.024501,2022arXiv220212178R}.
In contrast, topological superconductors with non-zero Chern number feature
in-gap modes in the presence of non-magnetic impurities, rendering
artificial topological superconductors vulnerable to disorder\cite{PhysRevB.93.075129,PhysRevB.94.115166,Prada2020,Andersen2020}. This weakness to disorder
stems from the fact that non-magnetic scattering has a pair breaking effect\cite{RevModPhys.78.373,PhysRevB.94.104501}
in unconventional superconductors due to the non-trivial sign structure of the
superconducting order\cite{PhysRevB.98.024501}.

It is first instructive to examine the gap of the moire system
as a function of the location of a strong impurity ($w=2t$) in a $9\times9$ supercell, 
as shown in Fig.~\ref{fig:strong}a. 
We focus here on the case with a periodic array of impurities in the system
following the moire pattern, with a single impurity per moire unit cell, and we keep the same parameters values as we have introduced in section \ref{sec:model}. Fig.~\ref{fig:strong}a shows the full gap of the moire pattern, for a single impurity
per moire unit cell located at each potential location.
The gap is measured in units of the topological gap for the topological
pristine system $\Delta^T$.
In particular, it is observed that the location of the impurity strongly impacts
the gap of the system\cite{2022arXiv220105045Z}. 
The resulting gap of the defective system can range from the pristine
value, observed for impurities at the exchange maximum, to nearly zero, observed
for impurities away from the exchange maximum (Fig.~\ref{fig:strong}a).
The previous phenomenology is directly reflected
in the electronic band
structure for different locations of the impurity (Fig.~\ref{fig:strong}bcd).
The electronic structure for an impurity halfway between
two exchange maxima is shown in Fig.~\ref{fig:strong}b, which in particular shows
a dramatically smaller gap than the pristine system.
In stark contrast, for two locations of the impurities close to the exchange maxima
as shown in Fig.~\ref{fig:strong}cd, the gap of the system remains nearly the
same as in the pristine system. In particular, the system retains its topologically
non-trivial Chern number $C=2$ with a nearly identical gap for the central impurities
shown in Fig.~\ref{fig:strong}cd, whereas the gap is drastically reduced for the
location of Fig.~\ref{fig:strong}b.

The previous phenomenology highlights that the location of the non-magnetic
impurity in the
moire pattern has a critical impact on the superconducting gap of the system.
Such strong spatial dependence is fully absent both
in non-topological moire superconductors due to
Anderson's theorem\cite{Anderson1959}, as well as in non-moire
artificial topological superconductors due to the equivalence of the sites.

It is interesting noting that in a moire superconductor there
is a large mismatch between the length scales of the Bloch states
and the impurity, in comparison with a uniform superconductor.
While there is certainly a mismatch of length scales,
our results show that the effect of impurities is comparable
both in the absence and presence of a moire.
From the low energy point of view, a local impurity in real space
can be considered like a delta function potential, which
creates scattering between all wavevectors in reciprocal space.
In the presence of a moire, the minibands span a small
portion of the original Brillouin zone of the material. Nonetheless, due to the mixing between all wavevectors
driven by a local impurity, its effect in the minibands
is comparable.

\section{Weak impurities in topological moire superconductors}
\label{sec:weak}

\begin{figure}[t!]
    \centering
   \includegraphics[width=\columnwidth]{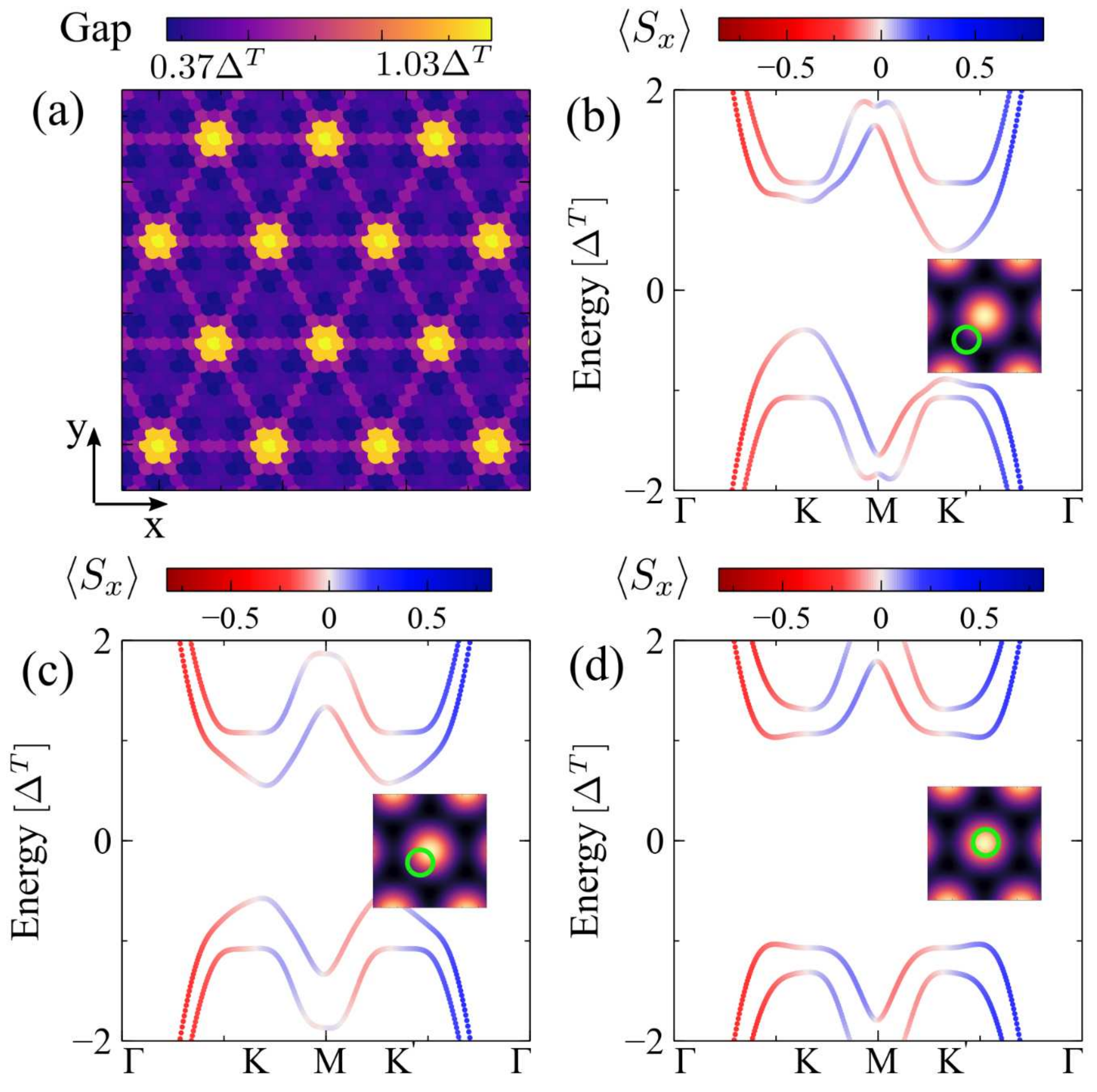}
    \caption{(a) Gap as a function of impurity location in the weak
    impurity regime $w=t/2$. Panels (bcd) show the electronic structure
    for three impurity locations, showing that halfway between
    exchange maxima, the gap is minimal
    (b). In contrast, close to the exchange maxima, the gap
    remains nearly unchanged (c,d).
    We used
    $J_{0}$=2$\Delta_{0}$, $\lambda$ =2$\Delta_{0}$, $\mu=3t$, $\delta_{J} = 2J_{0}$, $\delta_\Delta=1.4\Delta_{0}$.
    }
    \label{fig:weak}
\end{figure}

In this section, we examine the electronic structure
of the moire modulated model when
a periodic array of weak impurities
is distributed in the unit cell, with a single
impurity per moire unit cell.
In particular, we examine the evolution of the gap as a function
of the impurity location, which highlights the fine interplay between the
moire length and atomic defect. 
It is worth noting that examining the gap for weak impurities
allows tracking small changes in the gap as a function of the
impurity location, guaranteeing that the system remains in a topological phase.
As noted above, weak impurities would correspond to chalcogen substitution, such as
S in NbSe$_2$, leading to a local potential smaller than the bandwidth of the
low energy states.

We first look at the map of the spectral gap as a function of impurity location.
We take a moire unit cell with the size of $9\times9$, as shown in Fig \ref{fig:weak}a. It is observed that similar to the strong impurity regime
of Sec.~\ref{sec:strong}, the gap remains maximal close
to the maximum of the exchange profile, becoming smaller
in the other locations of the moire unit cell.
In comparison with the strong impurity limit, the weak impurity
allows keeping a sizable topological
gap even for the most detrimental locations away from the exchange maxima.
The previous phenomenology can also be observed by examining the
electronic structure for different locations of the impurity, shown in
Figs. \ref{fig:weak}bcd. In particular, it is observed that the electronic structure
remains similar for the three impurity locations, apart from small rearrangements
that account for the reduced topological gap.

\begin{figure}[t!]
    \centering
   \includegraphics[width=\linewidth]{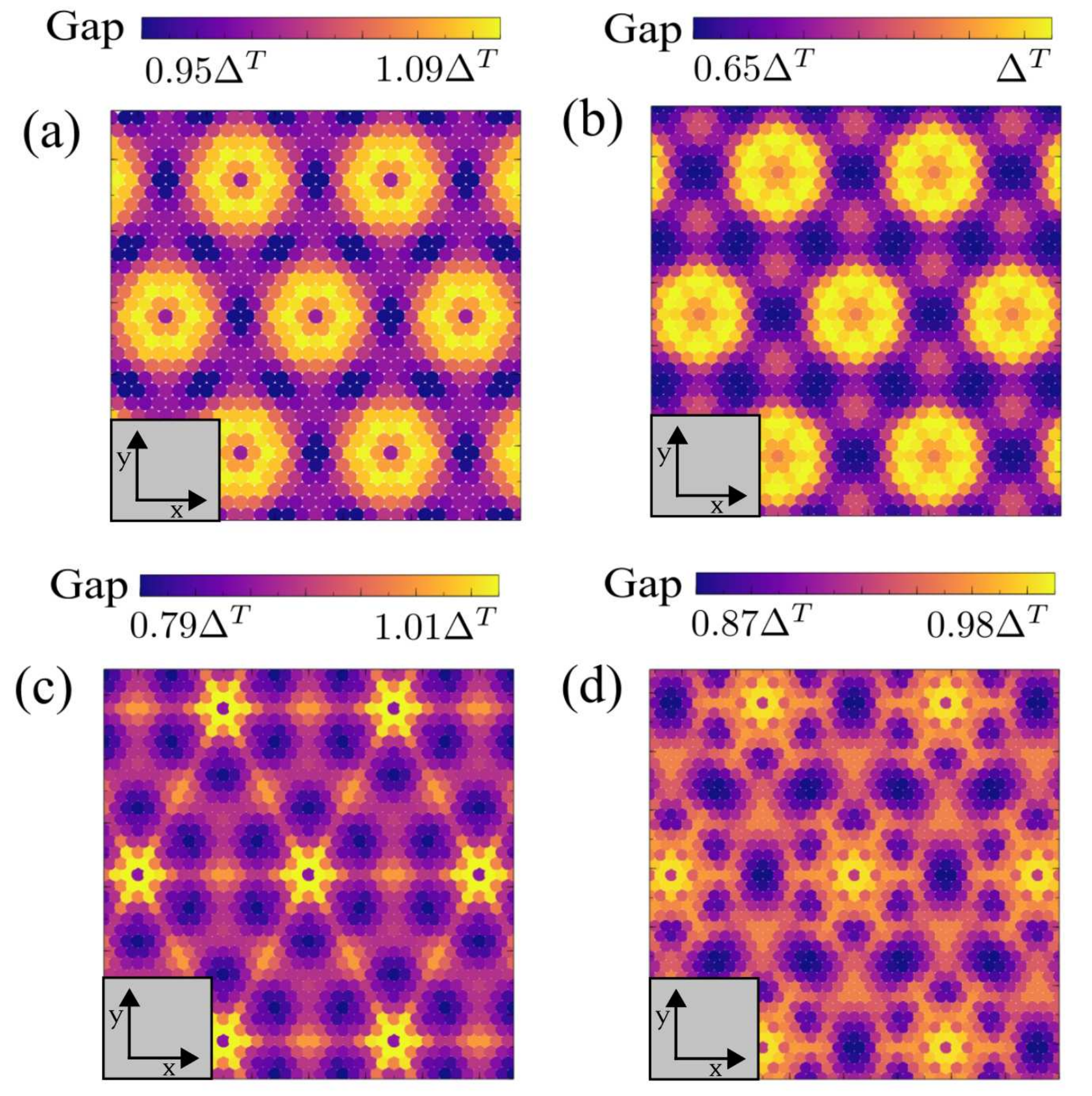}
    \caption{Topological gap as a function of the
    location of an impurity in a the moire unit cell, for 
    (a) 11$\times$11,
    (b) 13$\times$13
    (c) 15$\times$15
    (d) 17$\times$17
    supercells.
    It is observed a direct correlation between
    the location of the impurity
    in the unit cell and the moire
    pattern, leading to drastic changes in the
    energy gap.
    We used
    $J_{0}$=2$\Delta_{0}$, $\lambda$ =2$\Delta_{0}$, $\mu=3t$, $\delta_{J} = 2J_{0}$, $\delta_\Delta=1.4\Delta_{0}$.
    }
    \label{fig:gmap}
\end{figure}

The previous spatial dependence for different impurity locations can be analyzed
as a function of the moire length. The gap
for a single impurity per moire unit cell,
computed for different moire
lengths is shown in Fig.~\ref{fig:gmap}, where we consider
$11\times 11$ (Fig.~\ref{fig:gmap}a),
$13\times 13$ (Fig.~\ref{fig:gmap}b),
$15\times 15$ (Fig.~\ref{fig:gmap}c) and
$17\times 17$ (Fig.~\ref{fig:gmap}d).
In particular, we observe that close to the
exchange maxima, the local impurity
has a relatively weak impact, with the exception of the
exact center.
Away from the exchange maxima, the impurity
shows some of the most sizable effects, having
also a complex dependence with the moire length.
This complex dependence naturally emerges from the interplay
of between the moire length and the spatial
dependence of the in-gap mode, and is intrinsic
to any non-magnetic impurity in a moire
topological superconductor.
In particular, the interference between the in-gap state
and the moire pattern will be further addressed in Sec.~\ref{sec:single},
where we will consider a single impurity in an otherwise pristine moire system.

In this section, we have focused on addressing the fine interplay between
a periodic array of impurities and the moire pattern. In particular, it is observed
a dramatic dependence of the spectral gap on the location of the impurity, directly
correlated with the underlying moire pattern. In contrast with the strong impurity
case, weak impurities will keep the topological gap unchanged, and in particular,
all the defective superconducting states of this section
show the pristine Chern number $C=2$. 

\section{Impact of the moire amplitude}
\label{sec:amp}

In a moire superconductor, the moire pattern is characterized
by the amplitude of the modulation and its average value.
In this section, we analyze in detail the effect of moire amplitude,
allowing us to interpolate from the uniform to the
modulated limit. In particular, in the uniform limit,
the location of the impurity in the unit cell must lead
to identical gaps. In contrast, as the moire pattern is switched
on, the gap in the presence of an impurity
will develop a strong dependence on its location.
To track the evolution with the moire pattern, we 

\begin{figure}[t!]
    \centering
   \includegraphics[width=\linewidth]{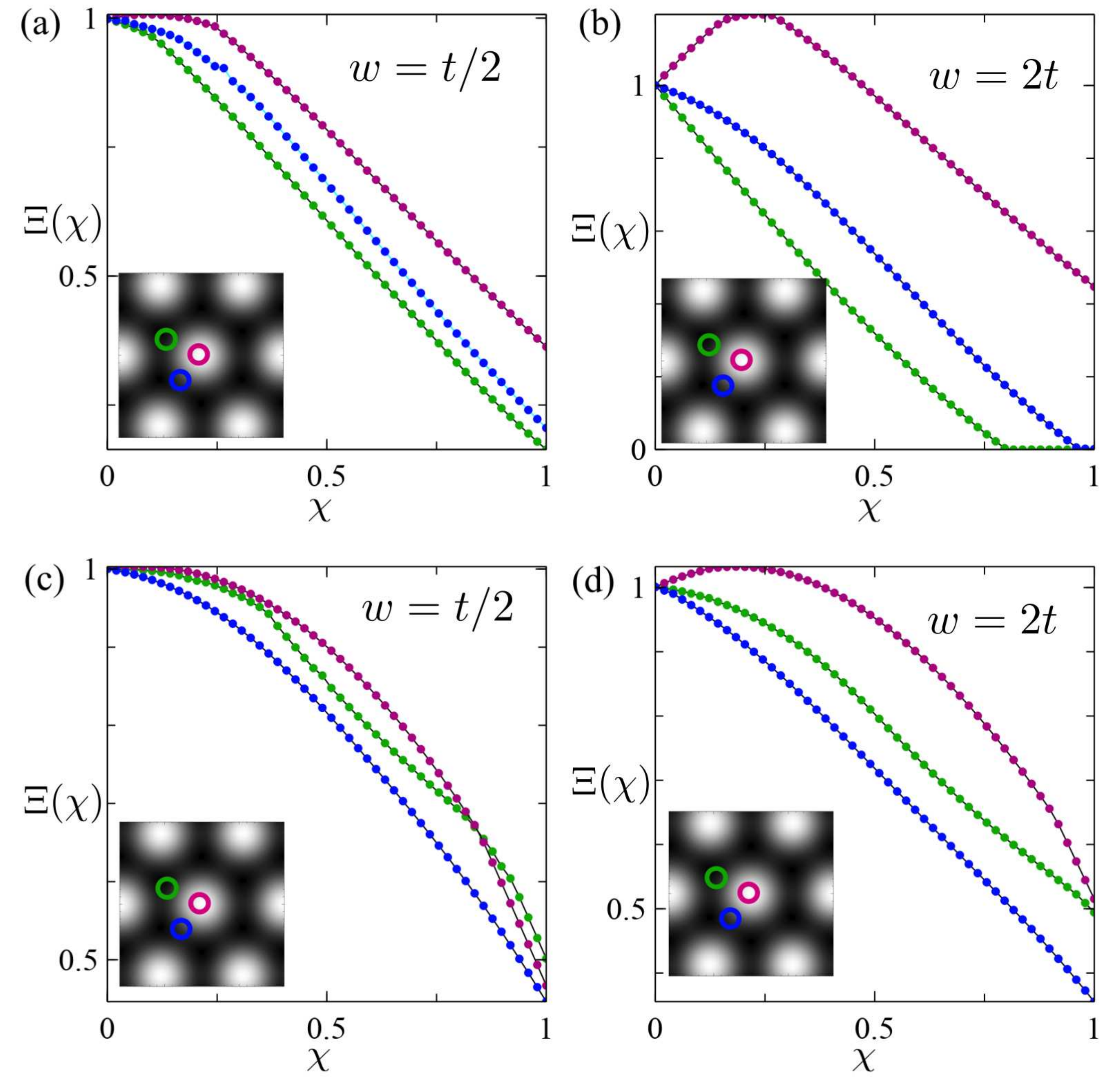}
    \caption{Normalized
    topological gap as a function of moire amplitude (Eq.~\ref{eq:normgap})
    for weak (a,c) and strong (b,d) impurity, for a
    $9\times 9$ (a,b) and a $11 \times 11$ (c,d)
    moire. It is observed that in the absence of
    the moire the location of the impurities does not impact the
    magnitude of the gap,
    whereas with moire, the topological
    gap shows a strong dependence on the size and location
    of the impurity.
    We used
    $J_{0}$=2$\Delta_{0}$, $\lambda$ =2$\Delta_{0}$, $\mu=3t$.
    }
    \label{fig:amp}
\end{figure}

keep the average values of the exchange and superconductivity constant, as well
as keeping constant ratios of their modulated amplitude and vary the $\chi$ parameter, defined in Eq. \ref{eq:amplitude}, between 0 and 1 ,which allows interpolating between the uniform and moire limit.

The topological gap for different locations of the impurity and moire unit cells
as a function of the moire amplitude is shown in Fig.~\ref{fig:amp}.
We consider two different moire unit cells, a $9\times 9$ moire unit cell (Fig.~\ref{fig:amp}ab) and a $11 \times 11$ moire unit cell (Fig.~\ref{fig:amp}cd), and two strengths
of the impurity $w=t/2$ (Fig.~\ref{fig:amp}ac) and $w=2t$ (Fig.~\ref{fig:amp}bd).
For different locations of the impurities, we compute the gap
of the moire system normalized to the gap of a uniform 
system with impurities. 

\begin{equation}
    \Xi (\chi) =
    \frac{\gap (\chi,w)}{\gap (\chi=0,w)}
    \label{eq:normgap}
\end{equation}

where $\gap$ is the gap of the system. The ratio $\Xi (\chi)$ allows to directly
observe the dependence on the location of the impurity
for different moire modulations.

\begin{figure*}[t!]
    \centering
    \includegraphics[width=\linewidth]{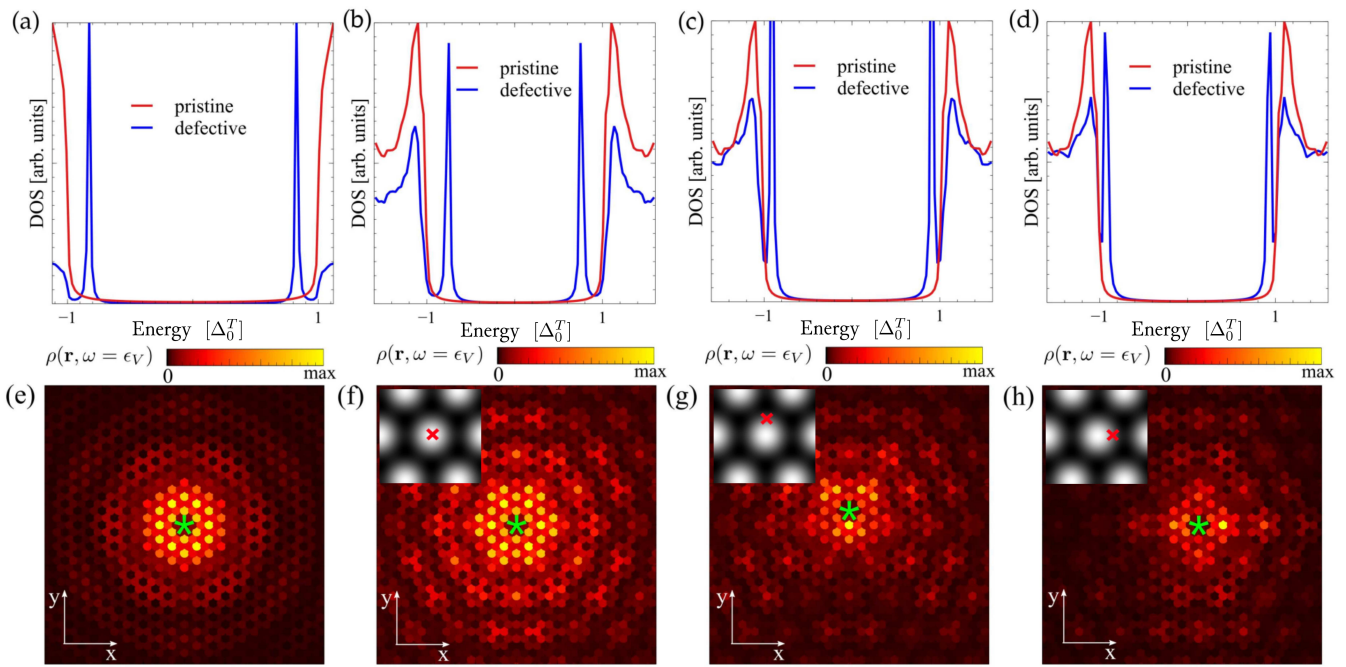}
    \caption{Density of states (a-d) and local density of states (e-h)
    at the energy of the ingap state $\epsilon_V$
    for a single impurity in an otherwise pristine system.
    Panels (a,e) correspond to the uniform case, whereas panels
    (b,c,d,f,g,h) to different locations of a single impurity (insets in f,g,h) for the moire case
    Both in the absence (a) and presence (b,c,d) of a moire, a strong non-magnetic
    impurity gives rise to an in-gap state. It is observed
    that in the presence of the moire pattern, the in-gap state
    leads to a strong interference with moire length, yielding
    a spatial dependence with respect to the location of the impurity
    (f,g,h).
    We used
    $J_{0}$=2$\Delta_{0}$, $\lambda$ =2$\Delta_{0}$, $\mu=3t$, $\delta_{J} = 2J_{0}$, $\delta_\Delta=1.4\Delta_{0}$ and moire $5\times 5$.
    }
    \label{fig:embbedding}
\end{figure*}

We now focus on the three locations of the impurities shown in Fig.~\ref{fig:amp},
one close to the exchange maximum of the moire (red), halfway between two exchange maxima (blue), and close to the center formed by three exchange maxima (green).
In the absence of a moire modulation, the three locations lead to the
same topological gap, while as the moire $\chi$ is turned on, the gap shows
a dependence on the location. In particular, we observe that in most of the
instances, the biggest
gap is obtained for an impurity close to the
exchange maxima (Fig.~\ref{fig:amp}), consistent with the results
obtained in Fig.~\ref{fig:gmap}. For the locations away from the exchange maxima,
the gap is maximized for different locations depending on the moire length, as
it is observed by comparing
Fig.~\ref{fig:amp}a with Fig.~\ref{fig:amp}c, and
Fig.~\ref{fig:amp}b with Fig.~\ref{fig:amp}d.
This phenomenology is also consistent with the moire dependence observed in
Fig.~\ref{fig:gmap}.
It is worth noting that,
as long as the gap remains open, the Chern number will
remain the same all the time. 
It is finally interesting to note that, for a strong impurity,
the location in the moire unit cell leads to substantially
bigger changes in the gap than a weak impurity.
This phenomenology is consistent with the results observed in
Fig.~\ref{fig:strong} and Fig.~\ref{fig:weak}.

The previous findings highlight that, in the presence of a moire modulation,
the location of an impurity
leads to different gaps, especially in the strong coupling limit.
Beyond the cases shown in Fig.~\ref{fig:gmap},
we note that even at the center of the unit cell, strong impurities
can give rise to a strong impact in the topological phase and ultimately lead
to a topological phase transition.
For other locations, the topological gap shows a complex interplay between
the location of the impurity and the moire length.
These results consider a single impurity per moire unit cell, leading
to strong overlap and interference between each in-gap mode. 

We finally comment on what would be the impact if disorder
is included in every single site.
In such limit, as the strength
of the disorder increases, the value of the
topological gap will decrease.
For weak disorder, the gap will remain finite,
yet smaller than the non-disordered case. 
However, for strong enough uniform
disorder, the system will effectively 
become gapless due to pair breaking effect of the
impurities.
This phenomenology is analogous to other 
unconventional non-swave superconductor, where non-magnetic disorder quenches
the underlying topological superconducting gap.

\section{Single impurity limit}
\label{sec:single}

In the sections above, we have focused on considering a periodic array
of impurities in the moire system. Here we will focus on a complimentary
limit, namely the case of a single impurity on an otherwise pristine moire system.
In this case, the moire system with a single impurity will lack any type of
translational symmetry, and therefore an electronic bandstructure
associated with a moire Bloch's theorem can not be
computed. In order to study the single impurity limit,
we will use Green's function embedding method, which allows to compute exactly
single defects in otherwise infinite pristine systems\cite{Lado2016,PhysRevResearch.2.033466}.

The embedding method relies on writing down the Dyson equation for the
defective system, that takes the form

\begin{equation}
    G_V(\omega) = [\omega - H_V - \Sigma(\omega)+i0^+]^{-1}
    \label{eq:gv}
\end{equation}

where $G_V(\omega)$ is the Green's function of the defective model, $H_V$ the Hamiltonian
of the defective unit cell, and $\Sigma(\omega)$ the selfenergy induced by the rest of
the pristine system. Solving the previous equation requires
deriving the selfenergy of the pristine system $\Sigma(\omega)$.
The selfenergy $\Sigma(\omega)$ can
be obtained by writing down the Dyson equation for the pristine model

\begin{equation}
    G_0(\omega) = [\omega - H_0 - \Sigma(\omega)+i0^+]^{-1}
    \label{eq:sigma}
\end{equation}

with $H_0$ the Hamiltonian of the pristine unit cell. We can now
take Bloch's representation
of the pristine unit cell Green's function

\begin{equation}
    G_0(\omega) = 
    \frac{1}{(2\pi)^2} \int [\omega- H_{\mathbf{k}} + i0^+]^{-1}
    d^2\mathbf{k}
    \label{eq:g0}
\end{equation}

where $H_{\mathbf{k}}$ is the Bloch's Hamiltonian. By obtaining $G_0$ from Eq.~\ref{eq:g0},
the selfenergy $\Sigma(\omega)$ can be obtained from Eq.~\ref{eq:sigma},
which in turn allows to obtain the Green's function of the defective unit cell from
Eq.~\ref{eq:gv}.

Using the previous methodology, we can extract both the total and local density of states
for a single defect in the moire topological superconductor. The local 
density of states $\rho (\mathbf x,\omega)$ and full density of states $A (\omega)$ 
are obtained as

\begin{equation}
\rho (\mathbf x,\omega) = -\frac{1}{\pi} \sum_{s,\tau}
\langle \mathbf x,s,\tau | 
\text{Im} (G_V (\omega)) |
\mathbf x,s,\tau \rangle
\end{equation}

and 
\begin{equation}
\rho(\omega) = -\frac{1}{\pi} \text{Tr} [
\text{Im} (G_V (\omega)) ]
\end{equation}

where $s$ runs over spin and $\tau$ over electron-hole sector. With the previous methodology,
we now compute the density of states with a single impurity for the topological
superconductor without moire pattern (Fig.~\ref{fig:embbedding}).

It is first instructive to consider the impurity in
the uniform topological superconductor, shown
in Figs. \ref{fig:embbedding}ae. In particular, it is observed that the
existence of a strong non-magnetic impurity ($w=2t$) gives rise to an in-gap state
(Fig.~\ref{fig:embbedding}a), and that the spatial profile of such in-gap mode
is localized around the impurity (Fig.~\ref{fig:embbedding}e) and features
intensity oscillations in space. These results in the uniform limit directly suggest
that the moire pattern will give rise to a rich interference pattern with the in-gap state,
ultimately responsible for the strong dependence of the location of the impurity
observed in previous sections.

We now move on to consider the case with a finite moire pattern
and single impurity,
whose density of states is shown in Figs. \ref{fig:embbedding}bcd and
the local density of states is shown in
Figs. \ref{fig:embbedding}fgh, for different location of the impurity with respect to the center of the moire pattern, shown by the insets in fgh. In particular, we consider three different locations of the
impurities. As shown in Fig.~\ref{fig:embbedding}bcd for all the locations of the impurities,
we observe in-gap modes at energies $\epsilon_V$. When computing the local density of states
associated with those in-gap modes $\epsilon_V$, 
shown in Figs. \ref{fig:embbedding}fgh, it is observed that
the interference pattern between the bound state and the moire pattern
leads to a strong dependence depending
on the location of the impurity\cite{Liebhaber2019}. 
Furthermore, it observed that
the in-gap mode spans over several moire unit cells, highlighting that for the periodic array
considered in the sections above the in-gap modes between different unit cells have a
strong overlap.

The previous results highlight that impurities in the moire pattern give rise
to in-gap states whose wavefunctions can potentially span over several unit cells,
and lead to strong interference effects with the moire modulation. The previous phenomenology
accounts for the strong dependence of the topological gap as a function of the impurity location
observed in Sec.~\ref{sec:weak} and Sec.~\ref{sec:strong}. 

\section{Edge states}
\label{sec:edge}

\begin{figure*}[t!]
    \centering
   \includegraphics[width=\textwidth]{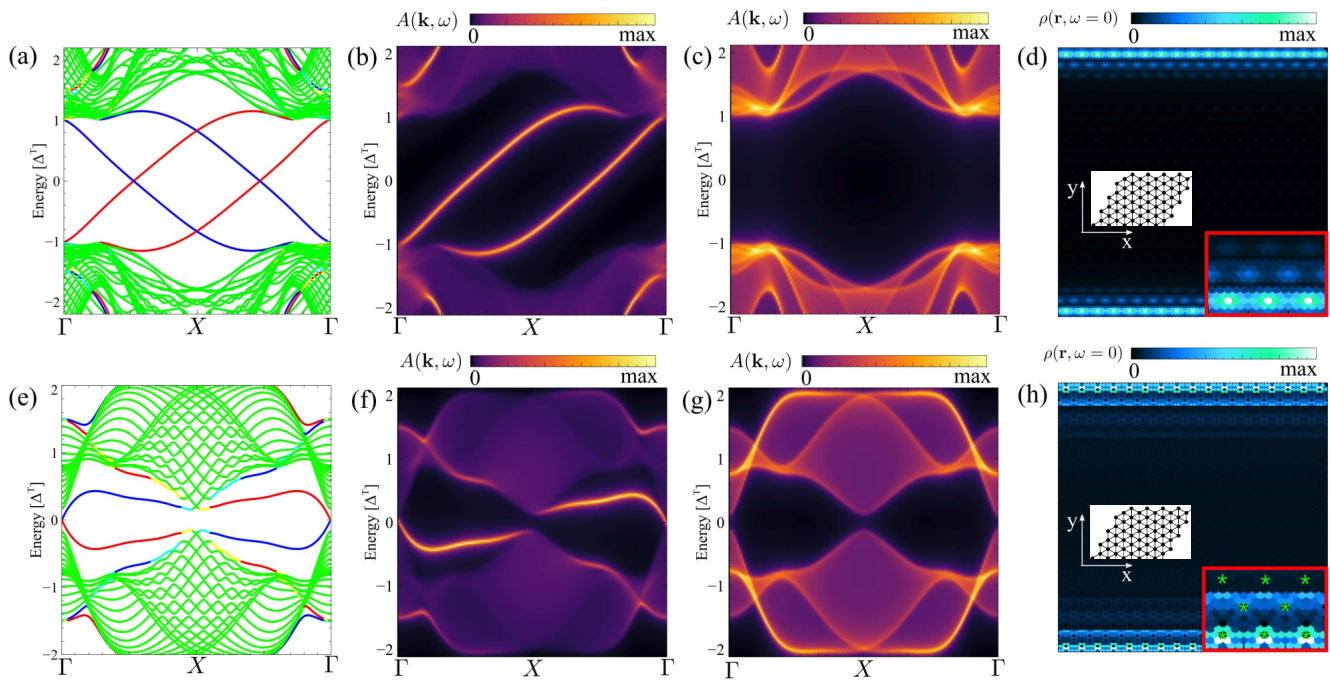}
    \caption{Moire topological superconducting states of a pristine system (a,b,c,d)
    and of a defective system (e,f,g,h).
    Panels (a,e) show the electronic structure of a ribbon infinite
    in the $x$ direction and finite in the $y$ direction.
    Panels (b,f) show the momentum-resolved edge spectral function,
    and panels
    (c,g) the momentum-resolved bulk spectral function.
    Panels (d,h) show the local density of states at $\omega=0$,
    highlighting the emergence of topological zero modes at the top and bottom
    edges following the moire pattern.
    The structure insets in (d,h) show a schematic
    of the boundary
    conditions used. 
        We used
    $J_{0}$=2$\Delta_{0}$, $\lambda$ =2$\Delta_{0}$, $\mu=3t$, $\delta_{J} = 2J_{0}$, $\delta_\Delta=1.4\Delta_{0}$, and moire $5\times 5$.
    }
    \label{fig:edge}
\end{figure*}

Finally, we analyze the emergence of edge states associated with the moire topological
superconducting state in a ribbon with the structure, inset in Fig. \ref{fig:edge}d,h, displaying a schematic of the boundary conditions, which is periodic along x direction and finite along y direction.

In particular, we will consider both pristine and defective cases,
and show that in both instances, the edge modes reflect the underlying moire pattern.
We first take a pristine system displaying the moire topological
superconducting state, as shown in Figs. \ref{fig:edge}abcd.
The electronic structure of a ribbon, infinite in the x-direction, is
shown in Fig.~\ref{fig:edge}a, displaying both the existence
of a gap in the bulk and propagating edge modes. Those two features
are more clearly seen by computing the edge (Fig.~\ref{fig:edge}b) 
and bulk (Fig.~\ref{fig:edge}c)
spectral functions as shown in Fig.~\ref{fig:edge}bc.
$A(\mathbf k,\omega)$ is the momentum-resolved spectral function in the surface and bulk of a semi-infinite ribbon,
with $\mathbf k$ the momentum in the translational invariant direction of the ribbon. It is
computed from the momentum-resolved Green's function
as
$A(\mathbf k, \omega) = -\frac{1}{\pi} \text{Tr} [
\text{Im} (G (\mathbf k,\omega)) ]$
where $G (\mathbf k,\omega)$ is computed with a
renormalization algorithm\cite{Sancho1985}.
In particular, it is observed that the edge hosts two co-propagating
modes (Fig.~\ref{fig:edge}b), consistent with the electronic structure
of the moire ribbon of Fig.~\ref{fig:edge}a. Beyond the existence
of edge modes, the moire pattern gives rise to a unique feature
in real space, namely the modulation of the edge modes
following the moire pattern. This can be clearly seen in
Fig.~\ref{fig:edge}d, where it is observed that the zero-energy modes
directly reflect the underlying moire pattern in the Hamiltonian.

We now move on to consider the defective system. In particular, we focus
on a moire superconductor with a single strong impurity per moire unit cell.
We emphasize that, depending on the location of the
impurity and the moire length, the topological phase can remain the same, 
become gapless, or a topological phase transition can take place.
For the sake of concreteness,
here we take a location of an impurity that
strongly disrupts the original
topological state, giving rise to a topological
phase transition to a phase with Chern number $C=-1$.
The electronic structure of the defective ribbon is shown in Fig.~\ref{fig:edge}e,
where it observed the existence of a small bulk gap and edge modes. Those modes
can be more clearly observed by computing the
edge (Fig.~\ref{fig:edge}f) and bulk (Fig.~\ref{fig:edge}g) spectral function as shown in 
Figs. \ref{fig:edge}fg. In particular, the edge spectral function now displays a single
edge mode as shown in Fig.~\ref{fig:edge}f, as expected from the bulk Chern number $C=-1$.
It is also observed that the topological edge states avoid the location of the impurity,
marked with a green star in Fig.~\ref{fig:edge}f, as expected from the strong impurity limit.
The edge modes reflect the moire periodicity again as shown in Fig.~\ref{fig:edge}h,
leading to the imprinting of the moire pattern in the topological edge modes.

The defective case considered above focuses on a periodic array of strong impurities. In real
experiments, impurities can appear either randomly distributed or can be engineered in arrays
using atomic manipulation.
The first case would correspond to chemical impurities intrinsically appearing during synthesis
of the material. In this situation, depending on the
density of impurities and their respective location, the original moire
topological phase will be disrupted, either by decreasing its topological gap or ultimately by
leading to a gapless state due to the proliferation of in-gap modes.
In the situation in which a periodic array of impurities is engineered by means
of atomic manipulation\cite{Hirjibehedin2006,Kalff2016,RevModPhys.91.041001,Drost2017,Slot2017,Yang2021,PhysRevB.103.205424,GonzlezHerrero2016,PhysRevResearch.2.043426,impKezilebieke2018,impKezilebieke2019}, 
specific arrangements as those considered
in Fig.~\ref{fig:edge}efgh can give rise to a topological state with different Chern number.
We have verified that solely by changing the location in which the impurity is deposited
in the moire unit cell, the resulting electronic structure could result in topological
phases with different Chern number, trivial phases, or even gapless phases.
The previous results highlight that atomic manipulation on top of moire
topological superconductors provides a new potential degree of freedom to engineer
tunable topological superconductors\cite{PhysRevB.100.075420}, 
by exploiting the interplay between the moire length
and local impurities\cite{Brihuega2017,Ramires2019,PhysRevResearch.2.033357,PhysRevMaterials.3.084003}.

We finally note that this analysis focuses on a minimal model that accounts for the physics of van der Waals
ferromagnet/superconductor heterostructures\cite{Kezilebieke2020,Kezilebieke2021,Kezilebieke2022}. 
From the quantum chemistry point of view, 
our model does not account for all the microscopic
parameters, but rather focuses on an effective model capturing the physics of this
family of heterostructures. In order to provide
a microscopically accurate description, calculations would need to be carried out with
Wannierization procedures based on first principles density functional 
methods\cite{PhysRevLett.121.266401}. 
In particular, these methodologies would account for the modulations in all the
Hamiltonian parameters, including spin-orbit coupling, hoppings, and onsite energies. Furthermore,
relaxation effects would be directly captured with these methodologies\cite{PhysRevLett.121.266401}. We note, however, that for the current system, first-principles density functional theory methodologies are beyond the computational capabilities,
in particular in the presence of spin-orbit coupling, due to the large number of atoms
in the unit cell for such a moire structure.
As a results, our discussions focus on a model Hamiltonian, yet without aiming to
reach chemical accuracy for NbSe$_2$/CrBr$_3$ heterostructures.

\section{Conclusion}
\label{sec:summ}

To summarize, here we addressed the interplay between local impurities and moire effects
in topological moire superconductors, as those realized in
CrBr$_3$/NbSe$_2$ heterostructures.
In particular, our results highlight that, in contrast 
with conventional artificial topological superconductors, the impact of impurities on a moire
system can give rise to radically different properties
depending on their location in the moire pattern.
For strong impurities, we observed that solely depending on the location
of the impurity in the moire pattern, the electronic structure can show as similar
topological gap as in the pristine limit, a nearly gapless state or
a topological phase transition to a topologically different state.
For weak impurities, we showed that the 
topological superconducting gap shows a dependence both on the location
of the impurity and the moire length,
yet maintaining its topological nature for all locations.
Furthermore, using an embedding formalism we addressed the impact of single
non-magnetic impurities in otherwise infinite pristine moire systems.
In particular, the absence of interference between impurities allows to clearly
identify that the moire modulation drastically impacts the spatial
profile of the in-gap mode created by the non-magnetic scatterer.
Ultimately, we showed that the moire modulation further emerges in the topological
edge modes of the topological superconductor, both in the pristine moire limit and
the defective limit.
Our results highlight the rich interplay between local impurities and topological
moire superconductors, and put forward engineered atomic impurities as a powerful and versatile
strategy to engineer artificial van der Waals moire topological
superconductors.

\textbf{Acknowledgements}:
We acknowledge the computational resources provided by
the Aalto Science-IT project,
and the financial support from the
Academy of Finland Projects No. 331342 and No. 336243,
and the Jane and Aatos Erkko Foundation.
We thank G. Chen, P. Liljeroth and 
S. Kezilebieke for useful discussions.

\bibliography{biblio}

\end{document}